\documentclass[aps,prb,amsmath,amssymb,amsfonts,twocolumn,nofootinbib]{revtex4}

\usepackage{graphicx}% Include figure files
\usepackage{dcolumn}% Align table columns on decimal point
\usepackage{bm}% bold math
\usepackage{amsmath}
\usepackage{amssymb}
\usepackage{color}

\usepackage{adjustbox}
\usepackage{multirow}

\usepackage{soul}

\makeatletter
\renewcommand*\env@matrix[1][*\c@MaxMatrixCols c]{%
  \hskip -\arraycolsep
  \let\@ifnextchar\new@ifnextchar
  \array{#1}}
\makeatother

\newcommand{\diag}{\mathrm{diag}}
\newcommand{\const}{\mathrm{const}}

\newcommand{\im}{\mathrm{Im}}

\newcommand{\be}{\begin{equation}}
\newcommand{\ee}{\end{equation}}

\newcommand{\bea}{\begin{eqnarray}}
\newcommand{\eea}{\end{eqnarray}}
\newcommand{\la}{\langle}
\newcommand{\ra}{\rangle}
\newcommand{\dg}{\dagger}
\newcommand{\td}{\tilde}
\newcommand{\dgc}{\ddagger}

\definecolor{orange}{rgb}{1,0.5,0}

\newcommand{\MM}{\color{blue}}

\begin{document}

 \title{Dissipation-induced first-order decoherence phase transition
in a non-interacting fermionic system}

\author{M. V. Medvedyeva}
\affiliation{Institute for Theoretical Physics, Georg-August-Universit{\"a}t G{\"o}ttingen,
Friedrich-Hund-Platz 1, D-37077 G{\"o}ttingen, Germany}
\author{M. T. \u{C}ubrovi\'{c}}
\affiliation{Institute for Theoretical Physics, Universit{\"a}t zu K{\"o}ln, Z{\"u}lpicher Str. 77, D-50937 K{\"o}ln, Germany}
\author{S. Kehrein}
\affiliation{Institute for Theoretical Physics, Georg-August-Universit{\"a}t G{\"o}ttingen,
Friedrich-Hund-Platz 1, D-37077 G{\"o}ttingen, Germany}

\begin{abstract}
We consider a dissipative tight-binding chain. The dissipation manifests as 
tunneling  into/out of the chain from/to a memoryless environment.
The evolution of the system is described by the Lindblad equation.
Already infinitesimally small dissipation along the chain induces a quantum
phase transition (QPT). This is a {\it decoherence QPT}: the reduced density matrix of a
subsystem (far from the ends of the chain) can be represented as the
tensor product of single-site density matrices.  We analyze the QPT in the thermodynamic limit
by looking at the entropy and the response function in the bulk. We also explore the properties of the 
boundaries of the chain close to the transition point and observe that the boundaries behave as if they undergo a second-order phase transition with power-law divergence of the correlation functions and response function. 
Disorder is known to localize one-dimensional systems,  but the coupling to the memoryless
environment pushes the system back into the delocalized state even in the
presence of disorder.
\end{abstract}

\maketitle
\section{Introduction}
Coupling to the environment can significantly change the properties of a quantum system. Intuitively, the presence of dissipation leads to a decrease of coherence in the system. It can induce various types of phase transitions.~\cite{RNM, Sachdev,Sachdev2, Sachdev3, Werner, Top, Zni10, Pro08,Prosen10}
 
The best known example of such a transition is exhibited by the spin-boson model: there is a critical value of the interaction between the two-level system and the bosonic environment which localizes the system.~\cite{SpinBoson} A more complicated example is the superconductor-metal transition in dissipative nanowires,~\cite{Sachdev,Werner} which can be modeled as a dissipative XY-spin chain, with a coupling to the bosonic bath at every site of the chain. It was shown both analytically and numerically~\cite{Sachdev, Sachdev2, Sachdev3} that the system experiences a universal second-order phase transition at the critical value of the coupling to the environment. 

These are examples in the presence of the bosonic bath. Realistically, especially in condensed matter systems the bath can be also fermionic.~\cite{fermbath} It is possible to describe it in a similar manner as the bosonic bath in the spin-boson model, i.e. using the Feynman-Vernon formalism. However, it is rather complicated to consider more than one or two sites in such a formulation. The problem is often simplified by studying a Lindblad-type equation.~\cite{Lindblad,breuer} It corresponds to a memoryless bath. Physically, it means that the quasiparticles in the bath are assumed to have a much smaller dynamical timescale comparing to the excitations in the system.  
Even the memoryless dissipation induces a novel behavior in the quantum systems. For example, 
dissipation along the system can lead to the algebraic decoherence in strongly interacting systems.~\cite{Muenc}

Phase transitions have been observed in the presence of a particle or energy flow in various spin chains.~\cite{NonQPT} For example, the equilibrium phase diagram of the transverse field Ising model has two phases: ordered and disordered, while in the presence of particle flow a new phase appears which carries a non-zero particle flux.~\cite{Rac97} In this phase the correlation functions show oscillations and decay with a power-law. 

The density matrix of the non-equilibrium steady state (NESS) of a non-interacting fermionic system is associated with an effective Hamiltonian.~\cite{Top} 
In this formalism, phase transitions can be observed directly from the spectrum of the effective Hamiltonian, which shows features absent in the closed system. For example, a topological phase transition has been found in a cold atomic system subjected to laser irradiation.~\cite{Top} 

Equilibrium phase transitions are characterized by discontinuous derivatives of the free energy:~\cite{LL5} the order of the transition is equal to the order of the first discontinuous derivative. In a non-equilibrium situation the free energy is not a well-defined statistical quantity. The partition function, on the other hand, remains well-defined also for a non-equilibrium system, as well as  entropy, which is given by the logarithm of the number of microstates.~\cite{textbook} 
Starting from the partition function or entropy we can define the (non-equilibrium) susceptibilities even though the free energy is ill-defined.~\cite{LL5} The susceptibility diverges at the transition point.~\cite{textbook2} For the second-order QPT the divergence is physical and detectable, while it is a delta-function-like divergence for a first-order transition. It means that in an infinite system undergoing a first order phase transition, when the divergence equals the Dirac delta-function, we can only observe the step (discontinuity) in susceptibility, while the (infinitely narrow) Dirac delta peak is not measurable.

\subsection{Short overview}

In this paper we study the fermionic chain connected to the memoryless bath at every site of the chain, hence we consider the Lindblad equation for non-interacting fermions.~\cite{Kosov,Medv,Prosen10,Prosen12} The ends of the chain are connected to non-interacting memoryless leads.\cite{Medv, Zni} The difference in chemical potential induces the particle flow in the system. We find a first-order QPT that separates the regimes of coherent and dissipative transport along the chain. The coherent state is characterized by the constant current along the chain, while in the dissipative state the current induced by the coupling to the reservoirs decays exponentially inside the chain. 
QPT between the two happens already at an infinitesimally small coupling to the environment, i.e. the critical coupling value is zero. The transition can be understood microscopically from the fact that the density matrix is decomposed into the tensor product of
one-site density matrices  in the bulk. The phenomenological reason for the transition is breaking of the time-reversal symmetry by the dissipation along the chain.
From the thermodynamic point of view, the transition is a consequence of the entropy-per-site jump. 
The bulk susceptibility also has a jump at the transition.
These facts make us conclude that it is a first order phase transiton.
We also detect the jump of the steady state current at the ends of the chain for  sufficiently long chains.  We can observe this non-equilibrium QPT in the spectrum of the effective Hamiltonian of the NESS: the gap present for zero dissipation along the chain closes in the presence of dissipation.  A non-equilibrium  QPT in the system coupled to the Markovian bath has also been observed in the XY spin chain\cite{Pro08, Prosen10} and in the XX spin chain.~\cite{Zni10}

The phase transitions are normally considered in the thermodynamic limit and the effects of the boundaries (finite size effects) are neglected (or, in numerical work, systematically eliminated e.g. by finite size scaling). When we discuss the transiton between the coherent transport through the chain and decoherent state induced by dissipation, we cannot neglect the effects of the boundaries, because the particle current is due to the injection of particles at the ends of the chain. Therefore, we study  the correlation functions and the electrical susceptibility  at the ends of the chain and observe power-law divergences.

We also consider the workings of dissipation in the presence of disorder. We find that any memoryless dissipation extended along the chain  destroys the localization by disorder. This result supports previous studies by the scattering matrix approach~\cite{scatt} and the Landauer-type approach with decoherence.~\cite{analyt}  
The phase transition to the dissipative state is universal and preserved in the presence of disorder. 

\section{Model and formalism}
Evolution of a system of non-interacting fermions coupled to the memoryless bath is described by the Lindblad equation:
\bea i\frac{d\rho}{d\tau}&=& [H,\rho] + \nonumber \\ 
&+&i \sum_{\mu,i/o}\left(2 \ell^{(i/o)}_\mu  \rho \ell^{\dg(i/o)}_\mu - \{\ell^{\dg(i/o)}_\mu \ell^{(i/o)}_\mu,\rho\} \right), \label{liouv}\eea
where $\ell_\mu$ are the Lindblad operators responsible for the coupling to the bath, $H$ is the tight-binding Hamiltonian for free fermions:
\be H=  \sum_{\{ij\}} t_{ij} \left( a^\dg_ia_j+ h.c \right)+ \sum_i U_i a^\dg_ia_i ,\label{ham}\ee
with $\{ij\}$ denoting the links between the sites,  $t_{ij}$ is the hopping amplitude between sites $i$ and $j$ and $U_i$ is an on-site potential. 
Further we are interested in a one-dimensional chain of the length $L$ coupled to the source and the drain at infinite bias voltage~\cite{Gurv} at the ends of the chain: $\ell_1^{(i)}=\sqrt{\Gamma^{(i)}}a_1^\dg$, $\ell^{(o)}_L=\sqrt{\Gamma^{(o)}}a_L$.  
The chain is connected to the memoryless bath  at every site, so the Lindblad operators are $\ell^{(i)}_{\mu}=\sqrt{d\Gamma_{\mu}^{(i)}}a_{\mu}^\dg$ for the sources and  $\ell^{(o)}_{\mu}=\sqrt{d\Gamma_{\mu}^{(o)}}a_{\mu}$ for the drains, $\mu=2,\ldots,L-1$.
 From now on in the text and in the plots the $\Gamma_\mu$ values are
measured in the units of the hopping $t$, or in the other words we put $t=1$.

\subsection{Solving the Lindblad equation}
The solution of the Lindblad equations for non-interacting fermions is notably simplified in the super-fermionic representation,~\cite{Kosov, Medv}  which is based on the doubling of the degrees of freedom as in thermofield theory. Here instead of solving a differential equation for the evolution of the $2^L\times 2^L$  density matrix, the calculations are done with the $2L\times 2L$ matrices. The observables of the NESS are computed directly. What is more, the full-counting statistics of the transport through the ends of the chain can be obtained by introducing the counting field which yields to the generating function.~\cite{Medv,Zni} 

We evaluate the current along the chain  by averaging the local current operator over the NESS:
\be \hat{j_k} =-i t (a_{k}^\dg a_{k+1}-a_{k+1}^\dg a_k) \label{current}.\ee
At the ends of the chain the current and the Fano factor are given by the derivatives of the generating function.

The Liouvillian for non-interacting fermions in the super-fermionic representation becomes quadratic after performing the particle-hole transformation,~\cite{Medv} as the Liouvillian becomes diagonal in the basis $\{f,f^\dgc,\td{f},\td{f}^\dgc\}$. The density matrix of the NESS is a vacuum for the operators $f$ and $\td{f}$. As there exists a linear relation between the initial basis $\{a,a^\dg,\td{a},\td{a}^\dg\}$ and the basis $\{f,f^\dgc,\td{f},\td{f}^\dgc\}$, the density matrix of the NESS is quadratic:
\be \rho_{NESS}=\frac{\exp(\mathcal{H}_{mn}a^\dg \td{a}^\dg) |00\ra_{a\td{a}}} { \langle I| \exp(\mathcal{H}_{mn}a^\dg \td{a}^\dg) |00\ra_{a\td{a}} },~~
\mathcal{H}_{jn}=\td{\kappa}_{ni}^{-1}\kappa_{ji}, \label{EffHam}
\ee
where the matrix $\tau$ is connected to the matrix of the eigenvectors $P$ of the transformation which diagonalizes the particle-hole transformed Liouvillian,~\cite{Medv} namely
$T=P^{-1}$, $\kappa_{kj}=T_{kj}$ and $\td{\kappa}_{kj}=T_{k+L,j}$ for $k,j=1,\ldots,L$.
Notice that $i\mathcal{H}$ is a Hermitian matrix as $\rho$ is Hermitian,  and $\langle I |$ is the left vacuum, $|I\rangle = \sum_n |nn\rangle_{a\td{a}}$,~\cite{Kosov} where by $n$ we denote the state in the $a$-basis. 
Therefore, $i\mathcal{H}$ can be considered as an effective Hamiltonian of the NESS.

\section{Dissipation-induced phase transition}

In this section we first observe the dissipation-induced phase transition in the transport properties at the ends of the chain and in the bulk and then we characterize the transition in the thermodynamic limit. Afterwards we discuss some specific aspects of the transition at the ends of the chain by studying the response and correlation functions close to the ends and reveal its microscopic nature. Finally, we study the influence of the dissipation on the phenomenon of delocalization in disordered systems.
 
\subsection{Observation of the transition}

\begin{figure}[tb!]
\begin{center}i
A\includegraphics[width=0.85\linewidth]{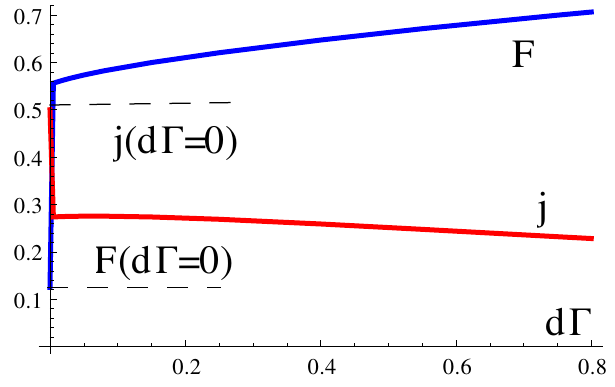}
B\includegraphics[width=0.85\linewidth]{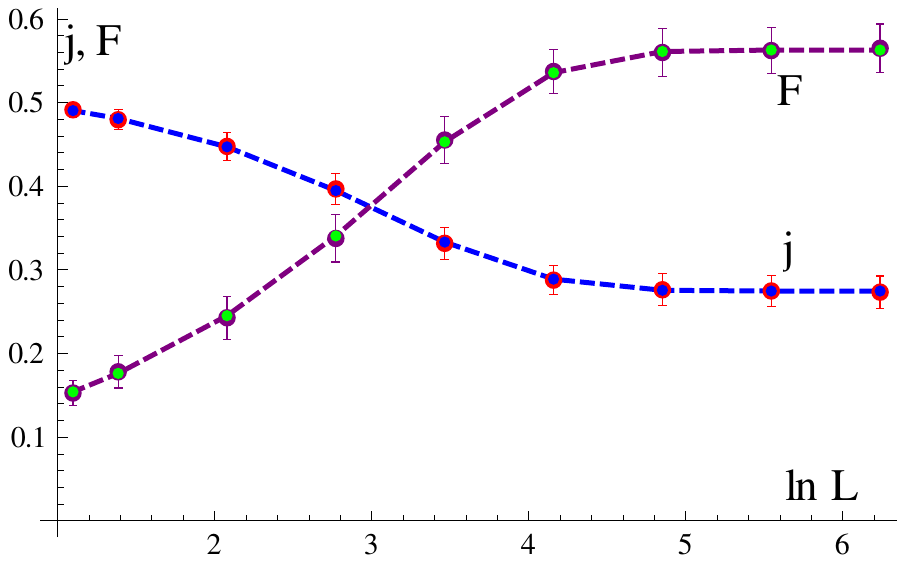}
\caption{\label{jump}
A: The jump of the current, $j$, and the Fano factor, $F$, at infinitesimally small dissipation  constant along the chain
$d\Gamma=d\Gamma^{(i)}=d\Gamma^{(o)}$ ($\Gamma^{(i)}=\Gamma^{(o)}=1$). 
B: Dependence of the current $j$ and the Fano factor $F$ through the ends of the chain on the length $L$ for random dissipation along the chain taken from the range 
$d\Gamma^{(i)},d\Gamma^{(o)}\in(0,0.04)$ (points with errorbars) and for the constant dissipation with the strength $d\Gamma^{(i)}=d\Gamma^{(o)} = 0.02$ 
(points and the dashed lines). 
Here and everywhere else in the text and the plots the $\Gamma_{\mu}$ values are measured in the units of the hopping $t$. 
}
\end{center}
\end{figure}

\begin{figure}[tb!]
\begin{center}
A\includegraphics[width=0.85\linewidth]{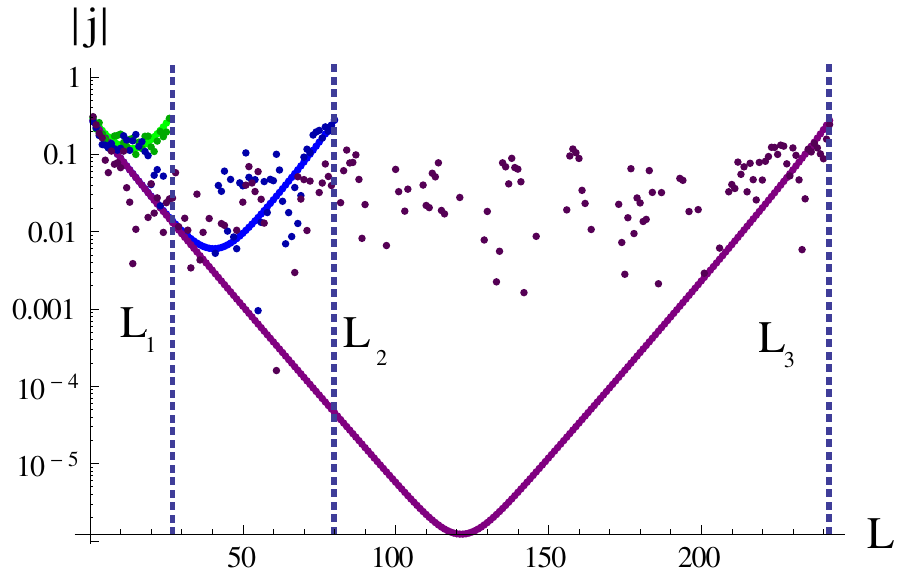}
B\includegraphics[width=0.85\linewidth]{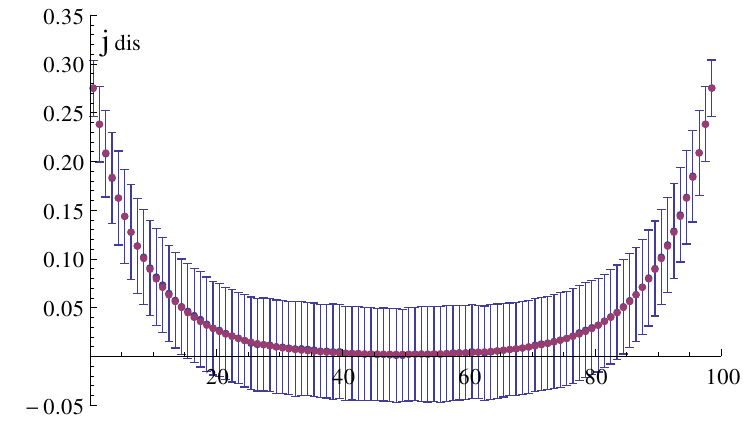}
\caption{\label{currentL}
Exponential decay of the current along the chain. A: Logarithmic scale, different lengths of the system. The currents in the system without randomness in dissipation are represented by the regular sets of points (forming solid lines). Darker, irregularly scattered points represent the current for one realization of the disorder in dissipation along the chain. B: The current through a dissipative chain after averaging over different disorder realizations. The scale is linear (not logarithmic) to show the standard deviation of the (fluctuating, random) current. Notice that the negative values of the current are physical, because some realization of the (random) couplings $d\Gamma$ can give ano overall current flowing in the opposite direction. 
The couplings at the ends of the chain are $\Gamma^{(i)}=\Gamma^{(0)}=1$, $d\Gamma=0.05$. For the average over disorder $d\Gamma^{(i)}_j,d\Gamma^{(o)}_j\in(0,0.1),~~j\in(2,L-1)$.
}
\end{center}
\end{figure}

\begin{figure}[tb!]
\begin{center}
\includegraphics[width=0.85\linewidth]{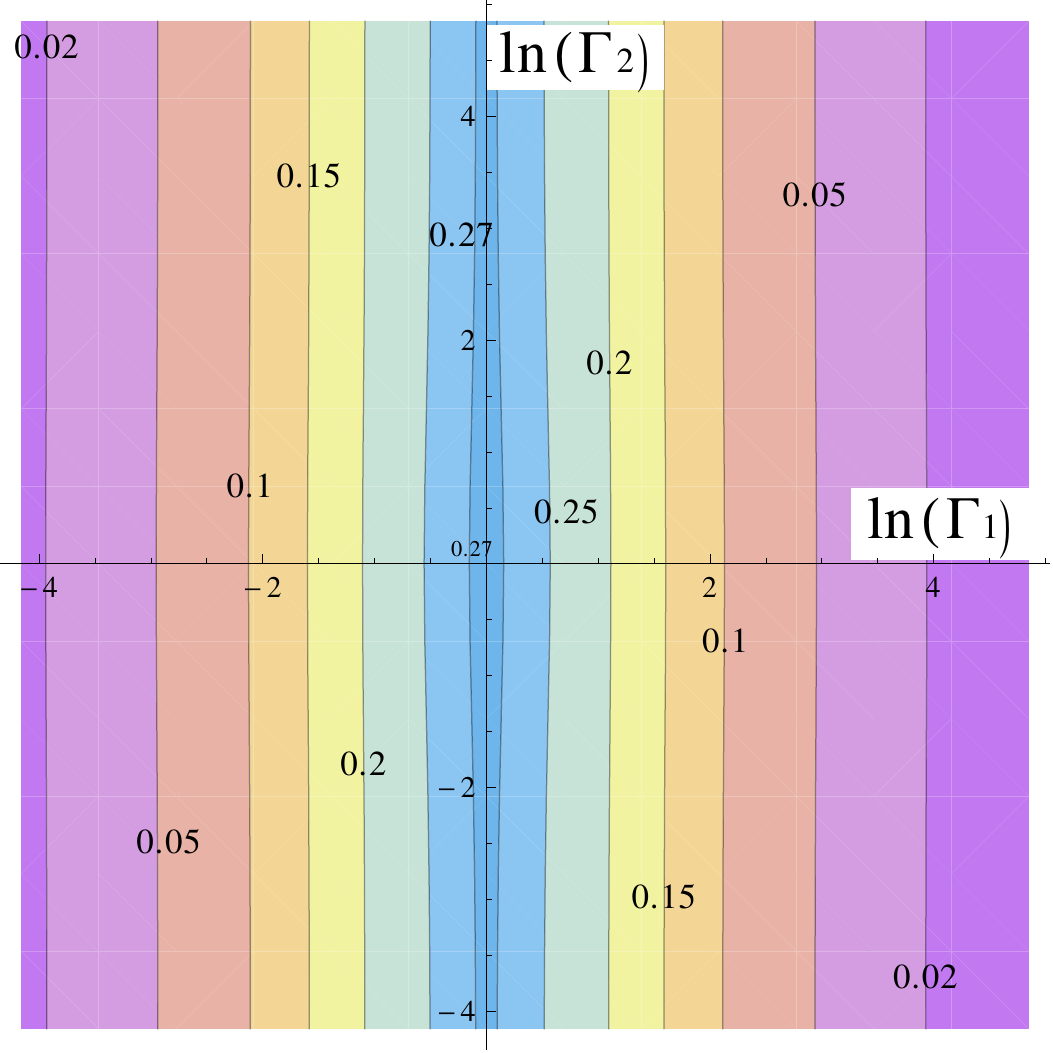}
\caption{ \label{dG}{\MM
Logarithmic plot of the current flowing from the system into the reservoir at the beginning of the chain (denoted by 1) as a function of the hopping rates at the ends of the chain, in the presence of the constant dissipation along the chain, $d\Gamma^{(i)}=d\Gamma^{(o)}=0.02$. 
Increasing the dissipation makes the current through one end independent of the coupling at the other end of the chain. In this plot we denote $\Gamma^{(i)}=\Gamma_1$, $\Gamma^{(o)}=\Gamma_2$.
}}
\end{center}
\end{figure}

We model dissipation along the chain as tunneling to the metallic gate in the absence of good isolation of the one-dimensional chain from the environment. To implement this we couple a source and a sink to every site of the chain.~\cite{Kosov}
We also allow for disorder in the hybridization strengths $d\Gamma_\mu^{(i/o)}$ to account for different tunneling rates to the environment. 

The fermionic chain coupled to the reservoirs only at its ends  has a uniform current along its length due to particle conservation.
Let us call the state of such a system coherent as the current at its ends depends on both couplings. On the other hand we  call the state of the system decoherent  when the current through a given end depends only on the coupling of the reservoir at this end.

We only expect to find a phase transition and the associated discontinuities in the thermodynamic limit, i.e. in an infinite system. For that reason we start by looking at a chain long enough that there is no dependence on its length, Fig.~\ref{jump}B. We see a jump both in the current and in the Fano factor when the dissipation is switched on, Fig.~\ref{jump}A.
Ref.~\onlinecite{Zni} provides the large deviation calculation for the current distribution function of the chain coupled to the reservoirs only at its ends. The current distribution is discontinuous as a function of the couplings to the reservoirs and the author suggests that this is the reason of the phase transition also for the system dissipative along its length.

In order to understand better the nature of the states on both sides of the transition, let us consider the current along the chain. We compute the expectation value of the local current operator~(\ref{current}) in the NESS for every link  of the chain. For a non-dissipative system it is constant along the chain due to the current conservation. For the dissipative case it decays exponentially inside the system, Fig.~\ref{currentL}.
One would certainly expect such behavior in the presence of the drains only. But in our setup we have both the source and the drain attached to every site of the chain. Therefore, we conclude that the exponential decrease of the current is connected to the coherence losses due to coupling to the memoryless environment, and not simply to the current leakage into the drains. 

If we allow for a random distribution of the dissipation along the chain, the current averaged over disorder configurations decays with the same exponent as the current in the system with uniform dissipation, with the magnitude equal to the mean of the distribution of the disordered couplings, Fig.~\ref{currentL}. 

With increasing dissipation strength, the current through one end of the chain becomes only weakly dependent on the coupling at the other end of the chain because the coherence of the transport through the chain is lost upon adding the dissipation along the chain,~Fig.~\ref{dG}.
Here we make a  plot for the constant dissipation rate along the chain since the current averaged over disorder in coupling strengths is the same as in the case of the constant dissipation (see Fig.~\ref{currentL}).

Both the presence of the jump in the transport characteristics at the ends of the chain and the coherence/decoherence transition in the current along the chain suggest that any non-zero dissipation along the chain induces the QPT.
It is not a van der Waals-type transition, meaning there is no analogue of the latent heat, that is, excitation of internal degrees of freedom, but the extra energy is instead exchanged with the bath.

\subsection{First-order phase transition in the thermodynamic limit}

Phase transitions are normally studied using the thermodynamic quantities and the response functions. In a non-equilibrium situation the partition function and the entropy are well-defined  thermodynamic quantities. Here we concentrate on the entropy and the response to the electric field, and eventually explain the microscopic nature of the transition.

\begin{figure}[tb!]
\begin{center}
\includegraphics[width=0.85\linewidth]{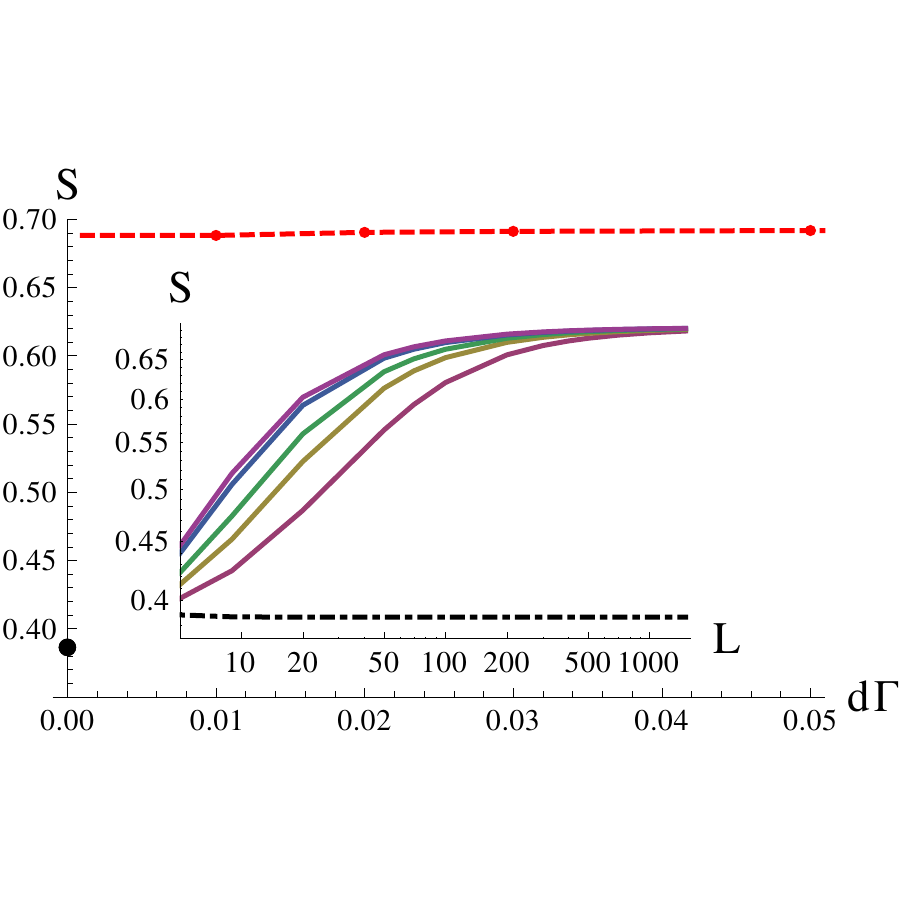}
\caption{ \label{Entropy}
Entropy jump at the transition point as a function of the dissipation strength. The dashed line is in agreement with Eq.~(\ref{SpecEnt}).
Inset: dependence of the entropy on the chain length for different dissipation strengths $d\Gamma=0.01,0.02,0.03,0.04,0.05,0.06$ (from top to bottom solid curve respectively), the dash-dotted line corresponds to the entropy in the absence of the coupling to the environment, the point at $d\Gamma=0$ at the main plot. 
}
\end{center}
\end{figure}

\subsubsection{Entropy}

The NESS is Gaussian, Eq.~\ref{EffHam}. Therefore, its effective Hamiltonian is a Hamiltonian of non-interacting fermions. In analogy with equilibrium statistical physics one can connect the entropy of the NESS to the eigenvalues $\mu_i$ of the effective Hamiltonian~(\ref{EffHam}):~\cite{Perchel}
\be \mathcal{S}= - \sum_i \left(\ln (1+e^{-\epsilon_i}) + \frac{\epsilon_i}{1+e^{\epsilon_i}}\right), \mu_i=e^{-\epsilon_i} \label{entropy}.\ee
The entropy per unit length $S=\mathcal{S}/L$ does not depend on the system length for sufficiently long systems and experiences a jump upon turning on the dissipation along the chain, Fig.~\ref{Entropy}.
For a chain without dissipation the specific entropy always depends on the couplings to the reservoirs  at the ends of the chain, while for a dissipative system it does not depend on the couplings to the leads in the thermodynamic limit (the contribution from the boundaries is of the order of $1/L$). The specific entropy tends to a value depending only on the ratio of the incoming and outgoing rates along the chain $\gamma=d\Gamma^{(i)}/d\Gamma^{(o)}$: 
\be S=\ln (1+\gamma)-\frac{\gamma}{1+\gamma}\ln \gamma.\label{SpecEnt}\ee
This corresponds to the entropy of the single site coupled to only two baths by the Lindblad operators $\sqrt{d\Gamma^{(i)}} a^\dg$ and $\sqrt{d\Gamma^{(o)}} a$. Indeed, the reduced density matrix of a site in the middle of the chain is the same as for a single site coupled to two baths up to a factor exponentially small in $L$. The coupling to the rest of the chain is irrelevant. The current in the middle of the chain vanishes, but what is happenning is even stronger: the correlation between two neigbouring sites vanishes exponentially  $\la c_{i+1}^\dg c_i \ra_{NESS} = O(\exp(-\beta i))$, where $i$ is the number of the site in the middle of the chain and $\beta$ is the slope of the exponential decay. Therefore, we can write down the reduced density matrix of the middle part of the system neglecting the exponentially small correlations between the sites as a tensor product of the density matrix of one site connected to two baths. 

\subsubsection{Spatial decoupling in the density matrix}
\label{subsubsec:TranslationalInv}
Such a spatial decoupling of a density matrix for a completely translationally invariant system (without current injection/removal at the ends) is evident. We can diagonalize the Liouvillian by the Fourier transform.
Indeed, in terms of Ref.~\onlinecite{Medv} the matrix $M$ after the Fourier transform obtains the block structure:
\bea \mathcal{L} &=&\sum_k (a_k^\dg~~ \td{a}_k) M_k
                    \begin{pmatrix}
                     a_k \\ \td{a}_k^\dg
                    \end{pmatrix} - i\sum_k (d\Gamma^{(i)}+d\Gamma^{(o)}),\\
                   M_k&=& \begin{pmatrix} 
                    - i\delta\Gamma +2t\cos k & 2d\Gamma^{(o)} \\ 
                      -2d\Gamma^{(i)}& i \delta\Gamma+2t\cos k 
                    \end{pmatrix}
\eea
with $\delta\Gamma=d\Gamma^{(i)}-d\Gamma^{(o)}$.
Each of the matrices $M_k$ can be diagonalized: $M_k= P_k^{-1}D_k P_k$, where $D_k$ is a diagonal matrix and $P_k$ is a matrix of eigenvectors. This transformation determines the basis where the Liouvillian is diagonal: 
\bea \begin{pmatrix}
     f_k \\ \td{f}_k^\dgc 
    \end{pmatrix} &=& P \begin{pmatrix}
     a_k \\ \td{a}_k^\dg 
    \end{pmatrix},~~~(f_k^\dgc~~\td{f}_k)=(a_k^\dg~~\td{a}_k)P^{-1}, \\
     \mathcal{L} &=& \sum_k \left(\lambda_k f_k^\dgc f_k-\lambda_k^* \td{f}_k^\dgc \td{f}_k \right).
\eea
Here we assumed that $D_k=\diag(\lambda_{k},\lambda^*_k)$ and $\im\lambda_k<0$. This structure leads to cancelation of the constant term in the Liovillian.

The steady state density matrix is determined as the vaccum of operators $f_k$ and $\td{f}_k$. The transformation to the basis of the $a,~a^\dg$ occupation numbers gives the density matrix:
\bea \rho &=& \sum_k \frac{\exp(\mathcal{H} a_k^\dg\td{a}_k^\dg )|00\rangle_{a_k \td{a}_k}}{{}_{a_k \td{a}_k}\langle I|\exp(\mathcal{H} a_k^\dg\td{a}_k^\dg )|00\rangle_{a_k \td{a}_k}}, \\  
\mathcal{H} &=& i\frac{d\Gamma^{(i)}}{d\Gamma^{(o)}},~~|I\rangle_{a_k \td{a}_k}=|00\rangle +|11\rangle
\eea
The effective Hamiltonian $\mathcal{H}$ is a constant, therefore the Fourier transform gives the density matrix which is a tensor product in position space:
\be \rho = \otimes_i \left(\frac{d\Gamma^{(o)}}{d\Gamma^{(o)}+d\Gamma^{(i)}} |00\rangle_{a_i\td{a}_i}+\frac{d\Gamma^{(i)}}{d\Gamma^{(o)}+d\Gamma^{(i)}}|11\rangle_{a_i\td{a}_i}\right).
\ee
We can thus conclude that  the density matrix is local in space. 
For the case of disordered leakage along the chain one cannot perform the Fourier transform of the Liouvillian analytically but numerical calculation shows that the density matrix averaged over disorder is again represented by the tensor product of single-site density matrices. For a single realization of the disorder in the couplings along the chain the decomposition is not exact, as shown in Fig.~\ref{currentL}A for the current through the chain for a single realization of the disorder.

\subsubsection{Response to the electric field}
\begin{figure}[tb!]
\begin{center}
\includegraphics[width=0.85\linewidth]{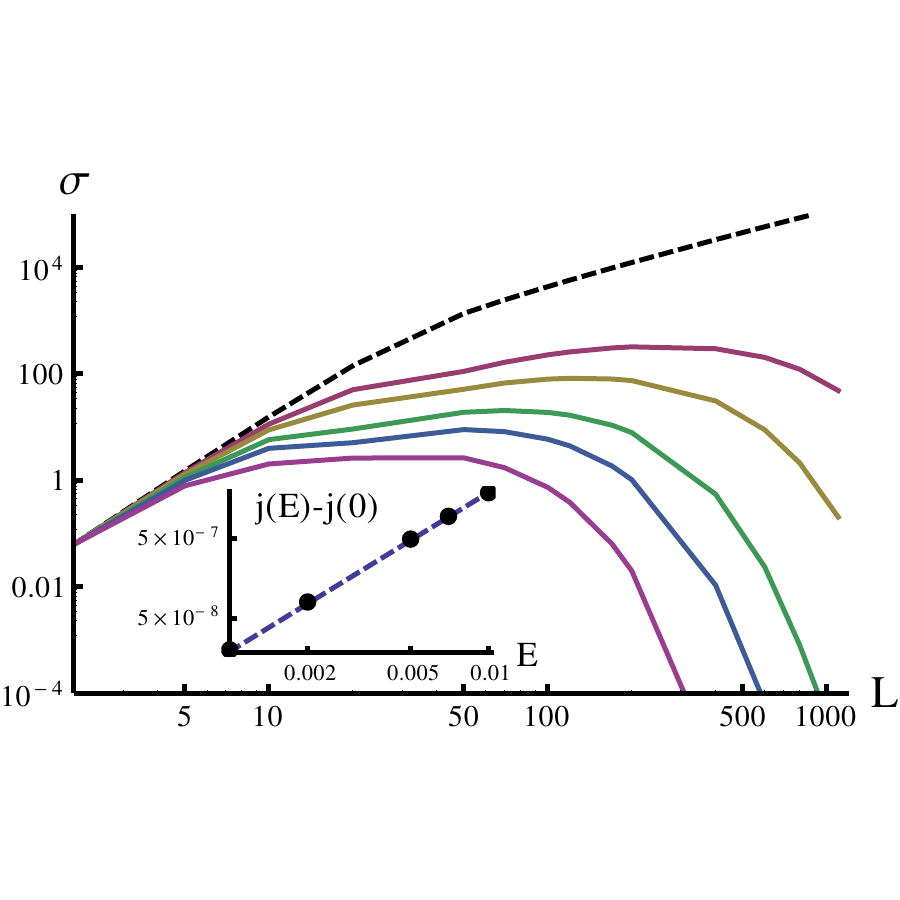}
\caption{ \label{CondBulk} 
Main plot: the convergence of the non-linear response to the electric field for long systems in the bulk of the chain. Solid lines correspond to different coupling strength $d\Gamma=0.005,0.01,0.02,0.03,0.05$ (from top to bottom) and the dashed line is  $d\Gamma=0$.
Inset: quadratic scaling of $j(E)-j(0)$ with the applied electric field (the scale in logarithmic).
}
\end{center}
\end{figure}

\begin{figure}[tb!]
\begin{center}
\includegraphics[width=0.85\linewidth]{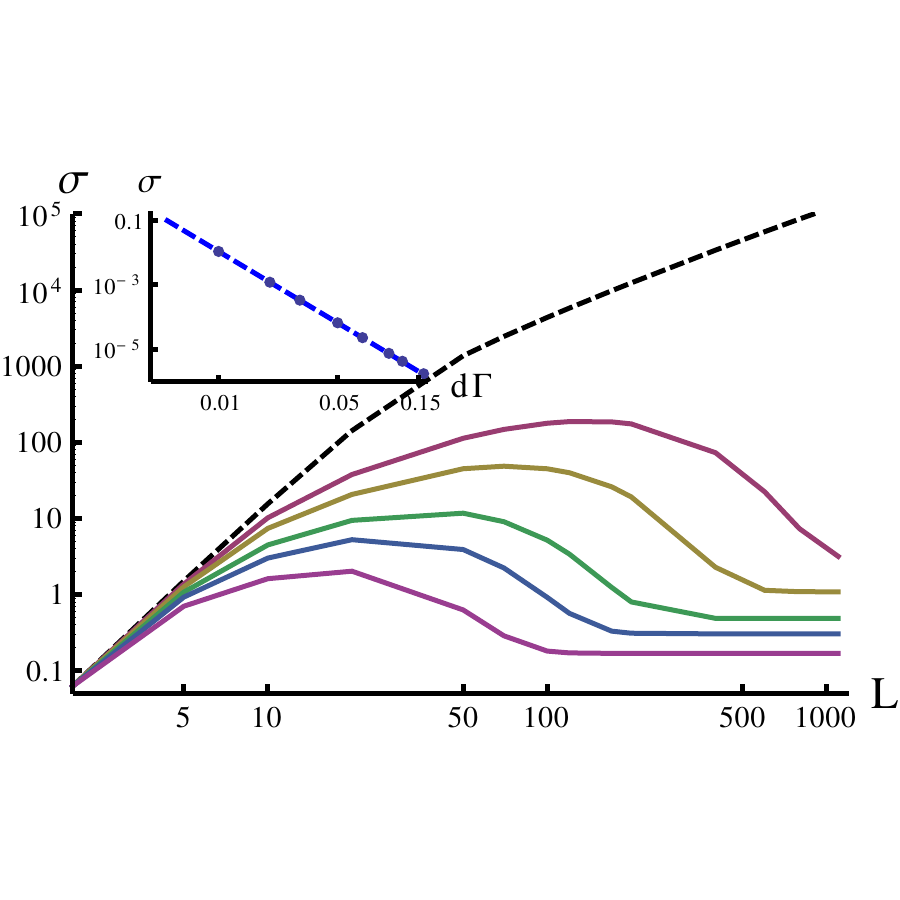}
\caption{ \label{CondBoundary} 
Main plot: the convergence of the non-linear response to the electric field for long systems at the ends of the chain. Solid lines correspond to different coupling strength $d\Gamma=0.005,0.01,0.02,0.03,0.05$ (from top to bottom) and the dashed line is  $d\Gamma=0$. Notice that the nonlinear conductivity at the ends points stays nonzero also in the thermodynamic limit. As in Fig.~\ref{CondBulk}, the conductivity is infinite in the absence of dissipation.
Inset: scaling of $\sigma$ with disorder strength with power-law fit: $\sigma = \alpha d\Gamma^\beta$, $\beta=3.161\pm0.001$.
}
\end{center}
\end{figure}

The response functions are good indicators of the equilibrium phase transitons. Let us consider a response of the current to a constant electric field $E$ applied along the chain. In the tight-binding model it is incorporated as a linearly growing on-site potential: $U_m = m E l_0$, where $l_0$ is the lattice constant.  In most models of the transport one assumes that the current flow is  due to an electric field applied along the system. Here we have a current through the chain due to the coupling to the reservoirs. The difference in on-site potential from site to site can be viewed as applying an additional field along the chain. For example, in a cold atom system one can imagine a lattice constructed with varying depths of the potential well. In the decoherent phase, the electric field changes the response function only locally: close to the ends we expect the susceptibility to be different from the middle of the system due to the presence of coherence because of the coupling to the reservoirs. 
The linear response of the current to the electric field applied along the chain  vanishes, and only the quadratic part is left, Fig.~\ref{CondBulk}, inset:
\bea j_{NESS}(E, d\Gamma^{(i)},d\Gamma^{(o)}; L)-j_{NESS}(0, d\Gamma^{(i)},d\Gamma^{(o)};L) = \nonumber\\
= \sigma(d\Gamma^{(i)},d\Gamma^{(o)};L) E^2. \label{cond}\eea
Here we also notice that there is a scaling with $E$: the dependence of the conductivity on length scales with $E^2$ for the same dissipation rates along the chain $d\Gamma^{(i)},d\Gamma^{(o)}$.
We attribute the quadratic dependence on $E$ to the structure of the NESS. 
The Ohm's law is an outcome of the linear response theory, which implies that the current is a consequence of the electric field applied to the equilibrium system. In our case the situation is tremendously different -- from the physical point of view, the current is already present in the system due to contact with the leads even before applying the electric field along the system. From the viewpoint of the response theory, the response is considered with respect to the non-equilibrium steady state. It is thus possible that the linear part of the response vanishes and only the non-linear part is present. 

The non-linear response to the electric field vanishes in the bulk of the chain, Fig.~\ref{CondBulk}. The response in the non-dissipative system grows infinitely in the thermodynamic limit because of the translational invariance in the bulk. Indeed, when we make the hopping parameters disordered (i.e. make them vary along the chain), the infinite growth of $\sigma$ is supressed. Therefore, there is a discontinuity in the value of $\sigma$ for infinitesimally small $d\Gamma$. It is consistent with the first order phase transition.

\subsection{Near-boundary effects}

\begin{figure}[tb!]
\begin{center}
A\includegraphics[width=0.85\linewidth]{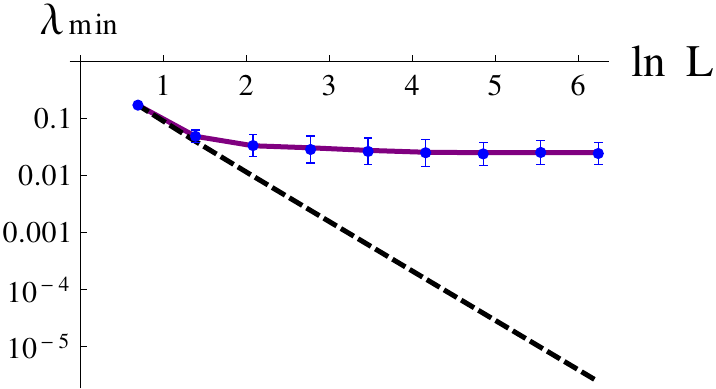}
B\includegraphics[width=0.85\linewidth]{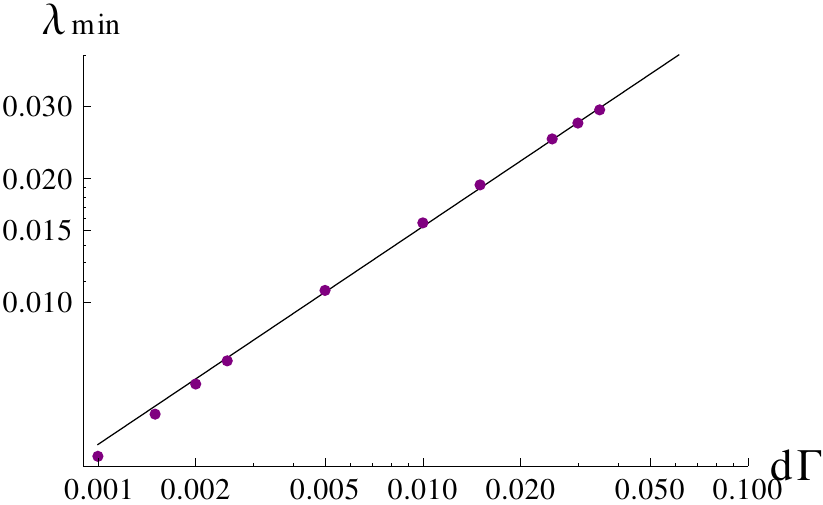}
\caption{\label{spectrum}
A: The lowest eigenvalue $\lambda_{min}$ of the effective Hamiltonian~(\ref{EffHam}) as a function of the system size for the system without (black dashed line) and with dissipation (blue points -- averages over the disorder from the range $(0,d\Gamma)$,  purple line -- constant dissipation with $d\Gamma/2$,  $d\Gamma=0.025$). 
B: The scaling of the lowest eigenvalue with disorder strength, $\lambda_{min}(d\Gamma)$, and the power law fit $\lambda_{min}\propto d\Gamma^\beta$ with $\beta=0.53\pm0.01$.  The couplings to the source and the drain are $\Gamma^{(i)}=\Gamma^{(o)}=1$.
}
\end{center}
\end{figure}

\begin{figure}[tb!]
\begin{center}
\includegraphics[width=0.85\linewidth]{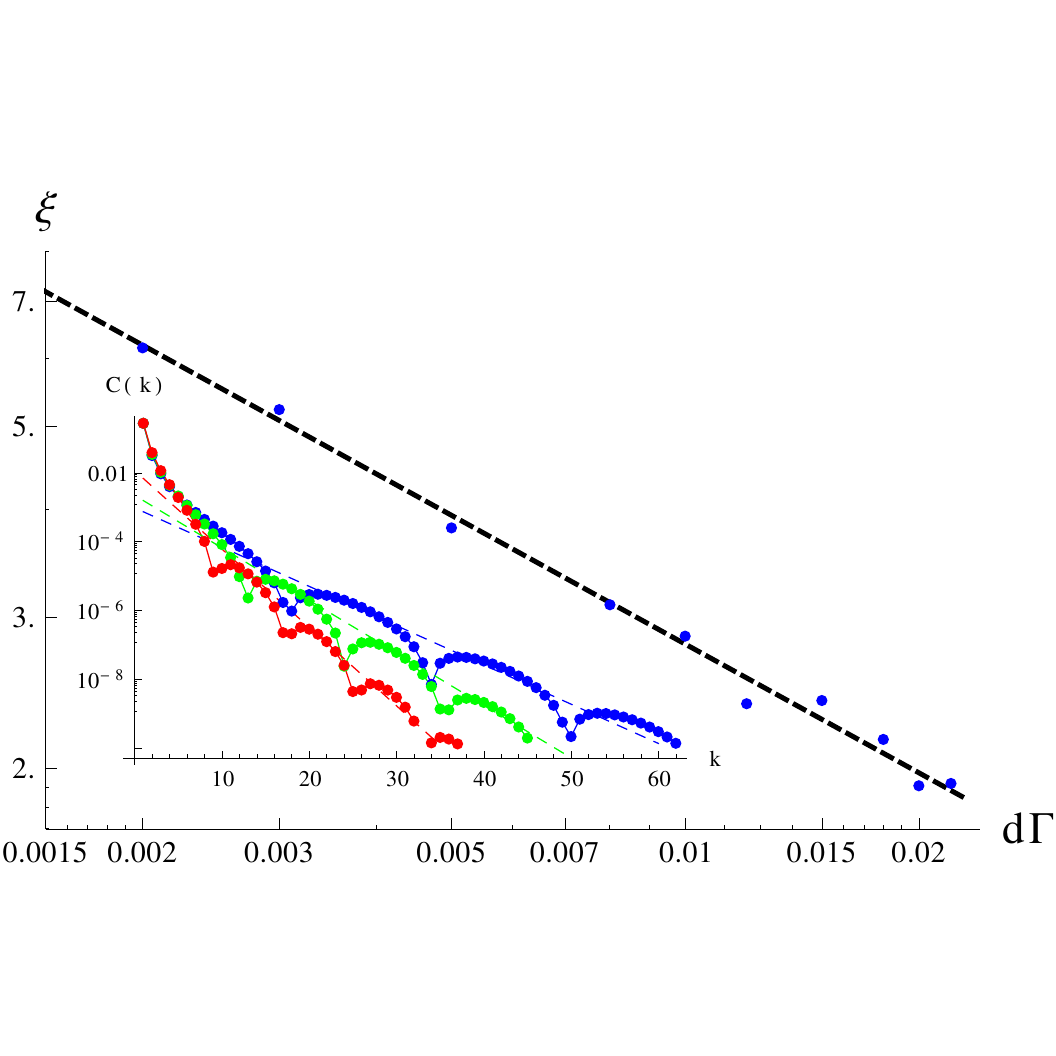}
\caption{ \label{Corr}
Dependence of the correlation length on the coupling to the environment for $\Gamma^{(i)}=\Gamma^{(o)}=1$ (dots) and the power law fit (dashed line). 
Inset: correlations at one end of the chain as a function the position for different couplings
strengths $d\Gamma=0.005,0.01,0.02$ (points) and exponential fits which determine the correlation length (dashed lines). 
}
\end{center}
\end{figure}

The symmetrized correlation function:
\be C_i(k)=\la a_{i+k}^\dg a_i+a_{i}^\dg a_{i+k}\ra_{NESS} \ee
provides further information about the transition. 
The correlations at the ends of the system are present and they  decay exponentially:  $C_i(k)\propto \exp(-k/\xi_i), i\sim1$ or $i\sim L$, where $\xi$ is a correlation length,~Fig.\ref{Corr}. We find the power-law divergence of the correlation length close to zero dissipation along the chain.
Inside infinitely long systems the correlations vanish: $\xi_i \rightarrow 0, i\sim L/2, L\rightarrow\infty$, as all coherence in the system is lost.

The non-linear condutivity converges to a non-zero value at the boundaries of the chain, Fig.~\ref{CondBoundary}, unlike in the bulk of the chain, where it converges to zero. This happens due to some remaining coherence at the ends of the chain. Even more, there is a power-law scaling of the conductivity with dissipation strength, the parameter which drives the phase transition, inset of Fig.~\ref{CondBoundary}.

To further corroborate the finding of the continuous QPT at the edges, let us now consider the spectrum of the effective Hamiltonian, $\mathcal{H}$. 
For the translationally invariant dissipative system from Sec.~\ref{subsubsec:TranslationalInv} the spectrum of the effective Hamiltonian is a delta-function  $\delta(\epsilon-\const \cdot d\Gamma^{(i)}/d\Gamma^{(o)})$, where the constant comes from the freedom of choice of the effective Hamiltonian, which is connected to the freedom of choice of constants in front of the left and the right vacuum of the Liouvillian. 
When we take into account the whole chain with the end sites, the spectrum of the effective Hamiltonian is influenced by the presence of the ends of the chain: in the absence of the dissipation along the chain the lowest eigenvalue $\lambda_{min}$ of $\mathcal{H}$ is $0$, while in the presence of the dissipation $\lambda_{min}$ shifts to a non-zero value, Fig.~\ref{spectrum}. There is a power-law scaling of $\lambda_{min}$ with the strength of the dissipation, Fig.~\ref{spectrum}B. 

\subsection{Disordered dissipative system}
\begin{figure}[tb!]
\begin{center}
\includegraphics[width=0.85\linewidth]{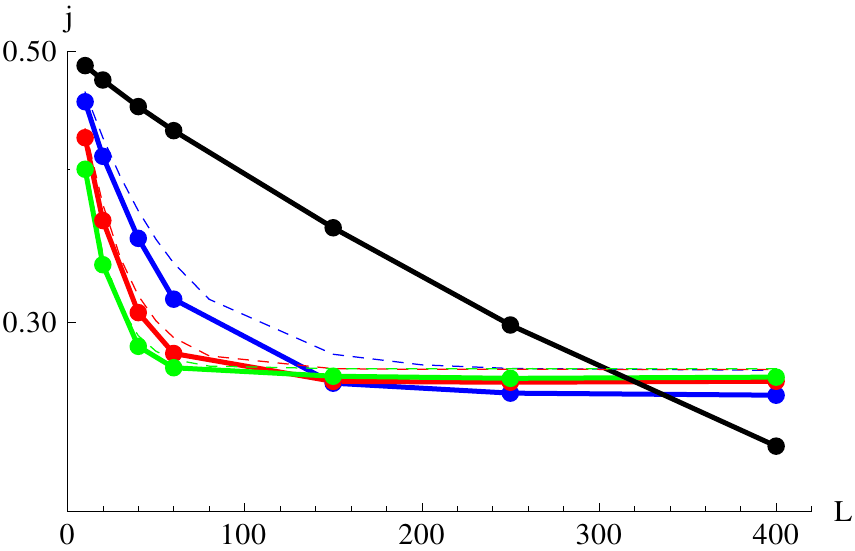}
\caption{\label{jL}
Dependence of the current through a disordered dissipative system on the length of the system for different values of the dissipation along the system, $d\Gamma=0,0.02,0.03,0.05$ (solid lines: $dU=0.3$, dashed lines: $dU=0$;$\Gamma_1=\Gamma_2=1$).
The current through the system is independent of the system length for a sufficiently long system.
}
\end{center}
\end{figure}

It is known that in one spatial dimension any disorder localizes the system.~\cite{Mirlin} 
However, the presence of dissipation changes this: {\it dissipation delocalizes the disordered system}, as the dissipation breaks the interference which is responsible for the localization. 

We have also checked the robustness of our finding to the influence of disorder, introducing on-site disorder taken from the uniform distribution with the range $(0, dU)$. 
The general phenomenology of the clean system with dissipation is preserved also in the disordered system. The current again reaches a finite (though smaller) value in the thermodynamic limit, and the current at one end only weakly depends on the coupling at the other end. An example is seen in Fig.~\ref{jL}, where for simplicity we consider constant couplings to the environment along the chain and average only over the disorder realizations of the on-site potential.  

\section{Conclusions and discussion}
We have considered the transport properties of a one-dimensional wire with leakage to the environment. In experimental systems, this leakage can happen due to misfabrication  and the presence of the tunneling from the wire to a metallic region underneath the wire.
We observe a first-order phase transition for infinitely long systems already at infinitesimal dissipation rate along the chain. From the microscopic point of view, this QPT means discontinuous behavior of the density matrix. On the macroscopic level it manifests itself in the jump in the current and the Fano factor. From the thermodynamic point of view we can say that the entropy jumps across the transition.
The specific entropy in the dissipative phase is equal to the entropy of a single site coupled to the source and the drain.

Essentially, the phase transtion is an anomaly: dissipation breaks the time reversal invariance. Upon taking the symmetry-breaking parameter (dissipation strength) to zero, we do not recover the result for unbroken symmetry.
In the continuum limit it is analogous to the fact
that, for example, viscosity effects in a fluid are non-perturbative and the flow undergoes a qualitative change for arbitrarily small nonzero dissipation: 
 the scaling exponents of the correlation functions of the velocities jump at the transition between an ideal and viscous liquid.~\cite{Falk}

In a different context, the transport theory for dissipative systems has been developed in Refs.~\onlinecite{scattG,scatt} in the language of the scattering matrices. 
Our Lindblad-based approach and the scattering approach are different in a few respects. 
First, let us consider a system without dissipation, coupled to two reservoirs at the ends. The scattering matrix theory describes the case when the wave coming from the reservoir into the system is coherent (just a plane wave), while the Lindblad approach describes the case of incoherent leads -- the hopping in the chain happens stochastically. This is also reflected in the transport properties: while for coherent transport the conductivity is proportional to the number of open channels in the system, for the transport induced by incoherent hopping it is not~\cite{Medv}.
Now let us move to the dissipative system. In the scattering matrix approach the dissipation is modeled through additional channels, which do not contribute to the transport (for one-dimensional non-dissipative problem the scattering matrix has the format $2\times2$,  for the incoming and the outgoing channel, while in the dissipative case the scattering matrix has a larger dimension, and only two channels  describe the transport along the chain whereas the others describe the scattering in the side channels). The dissipation constructed in this way is coherent, while the Lindblad-like dissipation is incoherent.

It is interesting that the spin system coupled to the bosonic bath at every site experiences a second order phase transition, and only at finite dissipation strength.~\cite{Sachdev,Sachdev2, Sachdev3, Werner} We do not know if the order of the transition is related to the presence or absence of  memory or it is determined by the statistics of the bath.

The phase transition in the quadratic fermionic systems was studied also in Ref.~\onlinecite{Pro08, Prosen10}. There, the XY chain coupled to the reservoirs at both ends was considered. The transition manifests itself in the change of behavior of the spin-spin correlation functions and the entanglement entropy, which does not depend on the system size on one side of the transition and grows linearly with the system length on the other side. The authors argue that the transition is of infinite order as all local observables are analytical across the transition. 
Subsequently the critical behavior has been observed also in the XX-spin chain~\cite{Zni10} coupled to the environment at every site of the chain: the correlation functions are short-ranged in the non-dissipative case, whereas they decay as a power-law in the presence of the on-site decoherence. 
The transition we observe is significantly different from the previously studied cases since it is of the first order. 
This probably happens because the Refs.~\onlinecite{Pro08, Prosen10} consider the local dissipation (only at the ends of the chain), while we are interested in the global dissipation. 
The difference with respect to the transition in the Ref.~\onlinecite{Zni10} lies in the fact that the NESS is not Gaussian: in our case the correlation functions in the presence of dissipation decay exponentially, while for the XX-chain with on-site dephasing there is a power-law decay of correlations.  

The current in the steady dissipative state of the system decays exponentially inside the chain, because the coupling to the environment decreases the coherence of the quantum system. 
For the random dissipation along the chain, we find that the average current decreases inside the system with the same exponent as for the chain with the same dissipation at every site which equals to the mean of the random coupling. 
One can try to measure   the current along the dissipative chain with a scanning tunneling microscope (STM): if it decreases  exponentially uniformly along the chain, then the dissipation model without disorder is a valid model, if the current inside the chain fluctuates, then the dissipation inside of the chain is random.  
The STM should be in the regime of a very low tunneling rate to the microscope tip, so that the tunneling to the tip does not destroy the dissipative state of the system itself.

We finish with an outlook. The state of the quantum system depends on the dimensionality, disorder, interaction, statistics, symmetries. 
The dissipation adds one more axis to the phase diagram.
It can lead to new types of behavior, already investigated in the spin-boson model~\cite{SpinBoson}, arrays of the dissipative Josephson junctions, dissipative spin chains.~\cite{Sachdev,Sachdev2,Sachdev3,Werner}
In the present paper we have investigated the behavior of the non-interacting fermionic system coupled to the Markovian bath and already have seen interesting quantum critical phenomena upon adding the dissipation along the chain. 
There are many unanswered questions: will this transition remain the first order upon adding memory to the bath, what happens to it in the presence of interactions, do dimensionality and symmetries influence the behavior of the dissipative system etc.

\section*{Acknowledgments}
This work was supported through SFB 1073 of the Deutsche Forschungsgemeinschaft (DFG). We are grateful to Achim Rosch and Fabian Biebl for careful reading of the manuscript and helpful discussions.


\begin{thebibliography}{99}

\bibitem{Zni10} M. \v{Z}nidari\v{c}, Stat. Mech. {\bf 2010}, L05002 (2010);  M. \v{Z}nidari\v{c}, Phys. Rev. E {\bf83}, 011108 (2011). 

\bibitem{Pro08} T. Prosen and I. Pi\v{z}orn, Phys. Rev. Lett. {101}, 105701 (2008).

\bibitem{Top} C.-E. Bardyn, M. A. Baranov, C. V. Kraus, E. Rico, A. Imamo\v{g}lu, P. Zoller and S. Diehl, New J. Phys. {\bf 15} 085001 (2013); 
O. Viyuela, A. Rivas, M. A. Martin-Delgado, Phys. Rev. B {\bf 86}, 155140 (2012).

\bibitem{RNM}  H. T. Mebrahtu, I. V. Borzenets, D. E. Liu, H. Zheng, Y. V. Bomze, A. I. Smirnov, H. U. Baranger and G. Finkelstein,  Nat. {\bf 488}, 61 (2012).

\bibitem{Prosen10} T. Prosen and B. Zunkovic, New. J. of Phys. {\bf 12}, 025016 (2010); 
T. Prosen, J. of Stat. Mech. P07020 (2010).

\bibitem{Sachdev} S. Sachdev, P. Werner and M. Troyer, Phys. Rev. Lett. {\bf 92}, 2370003 (2004).

\bibitem{Werner} P. Werner, {\it Dissipative Quantum phase transitions}, PhD thesis, ETH Zuerich (2005).


\bibitem{Sachdev2} S. Pankov, S. Florens, A. Georges, G. Kotliar and S. Sachdev, Phys. Rev. B {\bf 69}, 054426 (2004).

\bibitem{Sachdev3} P. Werner, M. Troyer and S. Sachdev, J. Phys. Soc. Jpn. Suppl. {\bf 74}, 67 (2005).


\bibitem{SpinBoson} A. J. Leggett, S. Chakravarty, A. T. Dorsey, M. P. A. Fisher, A. Garg and W. Zwerger, Rev. Mod.  Phys. {\bf 59}, 1 (1987).

\bibitem{fermbath} J. Jin, M. W.-Y. Tu, W.-M. Z, Y. J. Yan, New. J. Phys. {\bf 12}, 083013 (2010).

\bibitem{Lindblad} G. Lindblad, Rep. Math. Phys. {\bf 10}, 393 (1976).
\bibitem{breuer} H.-P. Breuer, F. Petruccione,  
{\it The Theory of Open Quantum Systems}, Oxford University Press, USA (2007).


\bibitem{Muenc} Z. Cai and T. Barthel, Phys. Rev. Lett. {\bf111}, 150403 (2013).

\bibitem{NonQPT} Z. Racz, {\it Nonequilibrium Phase Transitions}, Lecture Notes, Les Houches, July 2002. 

\bibitem{Rac97} T. Antal, Z. Racz, L. Sasvari,  Phys. Rev. Lett. {\bf 78}, 167 (1997).


\bibitem{LL5} L. D. Landau, E. M. Lifshitz, {\it Statistical physics}.

\bibitem{textbook} M. Le Bellac, F. Mortessagne and G. G. Batrouni, {\it Equilibrium and non-equilibrium statistical thermodynamics}, Cambridge Un. Press, New York, 2004.

\bibitem{textbook2} N. Goldenfeld, {\it Lectures on Phase Transitions and the Renormalization Group}, Perseus Books Publ., 1992.

\bibitem{Prosen12} S. Ajisaka, F. Barra, C. Mejía-Monasterio, and Tomaž Prosen
Phys. Rev. B {\bf 86}, 125111 (2012).
\bibitem{Kosov} A. A. Dzhioev, D. S. Kosov,  J. of Chem. Phys.  {\bf 134}, 044121 (2011); 
A. A. Dzhioev, D. S. Kosov J. Phys.: Condens. Matter {\bf 24}, 225304 (2012).


\bibitem{Medv} M. V. Medvedyeva, S. Kehrein, arxiv:1310.4997.

\bibitem{Zni} M. \v{Z}nidari\v{c}, Phys. Rev. Lett. {\bf 112}, 040602 (2014). 

\bibitem{Gurv} S. A. Gurvitz and Ya. S. Prager, Phys. Rev. B. {\bf 53}, 15932 (1996); S. A. Gurvitz, Phys. Rev. B. {\bf 57}, 6602 (1998).


\bibitem{scattG} K. Maschke, M. Schreiber, Phys. Rev. B {\bf 44}, 3835 (1991); 
G. Burmeister, K. Maschke, M. Schreiber, Phys. Rev. B {\bf 47}, 7095 (1993).

\bibitem{scatt} K. Maschke, M. Schreiber, Phys. Rev. B {\bf 49}, 2295 (1994).

\bibitem{analyt} D. Roy and N. Kumar, Phys. Rev. B {\bf 77}, 064201 (2008).

\bibitem{Perchel} I. Perschel and V. Eisler, J. Phys. A: Math. Theor. {\bf 42}, 504003 (2009). 

\bibitem{Mirlin} F. Evers and A. D. Mirlin, Rev. Mod. Phys. {\bf 80}, 1355 (2008).

\bibitem{Falk} G. Falkovich, K. Gawedzki, M. Vergassola, Rev. Mod. Phys. {\bf 73}, 913 (2001); G. Falkovich, J. Phys. A: MathTheor {\bf 42}, 123001 (2009).

\end{thebibliography}
\end{document}